\documentclass[prd,a4paper,nofootinbib,
showpacs, 
twocolumn,
]{revtex4}
\def\eq#1{{eq.~(\ref{#1})}}
\def\eqs#1#2{{eqs.~(\ref{#1})--(\ref{#2})}}

\def\vev#1{\left\langle #1\right\rangle}

\def\Tr{\mbox{Tr}\,}

\def\gtap{\ \raisebox{-.4ex}{\rlap{$\sim$}} \raisebox{.4ex}{$>$}\ }

\def\hbar{\hspace{0pt}\raisebox{1pt}{$-$} \hspace{-7pt} h}

\def\5{\overline 5}

\newcommand{\be}{\begin{equation}}
\newcommand{\ee}{\end{equation}}
\newcommand{\bea}{\begin{eqnarray}}
\newcommand{\eea}{\end{eqnarray}}
\newcommand{\nn}{\nonumber}

\begin{document}
\title[Flavor and electroweak symmetry breaking]{A heavy Higgs boson from flavor and electroweak symmetry breaking unification}
\date{\today
}
\author{F.~Bazzocchi}
\author{M.~Fabbrichesi}
\affiliation{INFN, Sezione di Trieste and\\
Scuola Internazionale Superiore di Studi Avanzati\\
via Beirut 4, I-34014 Trieste, Italy}
\begin{abstract}

\noindent We present a unified picture of flavor and electroweak  symmetry breaking based on a nonlinear sigma model spontaneously broken at the TeV scale. Flavor and Higgs bosons arise as  pseudo-Goldstone modes. Explicit collective symmetry  breaking yields stable vacuum expectation values and masses protected at one loop by the little-Higgs mechanism.  The coupling to the fermions generates well-definite mass textures---according to a $U(1)$ global flavor symmetry---that correctly reproduce the mass hierarchies and mixings of  quarks and leptons. The model is more constrained than usual little-Higgs models because of bounds on weak and flavor physics. The main experimental signatures testable at the LHC are a rather large mass $m_{h^0} = 317\pm 80$ GeV for the (lightest) Higgs  boson   and a characteristic spectrum of new  bosons and fermions at the TeV scale.
\end{abstract}
\maketitle
%
\textbf{Motivations.}
Naturalness of the standard model  seems to demand new physics at or around the 1 TeV scale. On the other hand, precision measurements do not show any departure from standard physics up to roughly 10 TeV. This \textit{little hierarchy} problem is solved in little-Higgs models~\cite{littlehiggs} by introducing new particles near 1 TeV in a sufficiently hidden fashion as not to show in the precision tests. 
 
 The Higgs sector of the standard model  also contains the physics of flavor in the Yukawa lagrangian, that is,  fermion masses and mixing matrices of the three generations. What happens if we try to include flavor (horizontal) symmetry breaking in the little-Higgs models? At first sight, this seems impossible because of the much higher scale of the order of $10^4$ TeV (mainly coming from bounds on flavor changing neutral currents) at which flavor symmetry breaking should take place. However, the bounds depend on the specific realization of the symmetry breaking and they are not necessarily so strong if the flavor symmetry is global and there are no flavor charged gauge bosons~\cite{littleflavons}. 
 
 As we  show in this Letter, it is indeed possible to take closely related breaking scales for  the electroweak and flavor symmetries and thus unify the two in a single little-Higgs model. 
  This unification, beside solving the little hierarchy problem provides---along the lines of the Froggart-Nielsen mechanism~\cite{flavormodels-old}---stable  textures of a well-defined type that correctly reproduce the mass hierarchies and mixings of  quarks and leptons. 
    
 The model gives a characteristic spectrum testable at the LHC of new particles, in addition to those of the standard model, and  a lightest Higgs boson mass that is  more constrained than in the usual, electroweak only, little-Higgs models. 
  
\textbf{The sigma model.}
 In order to have a  single, unified model \textit{\`a la} little Higgs describing  the entire flavor structure as well as the electroweak symmetry breaking,  the Higgs boson and the flavons must be  the pseudo-Goldstone bosons of the same spontaneously broken global symmetry.
These pseudo-Goldstone bosons---we shall call them \textit{flhiggs}---should transform under both  flavor  and    electroweak symmetries.
 
 We consider  a  $SU(10)$ global symmetry  spontaneously broken to
$SO(10)$ at the scale $f$.
Fifty-four generators of  $SU(10)$ are
broken giving $54$ real Goldstone bosons we parametrize in a non-linear sigma model fashion as
\be
\Sigma (x) = \exp \left[ i \Pi (x) /f \right]
\Sigma_0 \, ,
\ee
with $\Pi(x)= t^a \pi^a(x)$, where  $t^a$ are
the broken generators of  $SU(10)$, $\pi^a(x)$ the fluctuations
around the vacuum $\Sigma_0$ given by
\be
\Sigma_0 \equiv \langle
\Sigma \rangle = \left( \begin{array}{cc|cc} 0 & I_{4\times 4}  & 0 & 0 \\ I_{4\times 4} & 0 & 0 & 0 \\
\hline
0 & 0 & 0 &1  \\ 0 & 0 &  1  & 0 \\
\end{array} \right) \, .
\label{vacuumsigma} \ee
Within $SU(10)$ we identify seven subgroups
\be
SU(10) \supset  U(1)_F \times [SU(3) \times U(1)]_W^2 \times [U(1)_X]^2\, ,
 \ee
where the  $U(1)_F$ is the global flavor symmetry while the $[SU(3) \times U(1)]_W^2$
are  two copies of an extended electroweak gauge symmetry. The standard model weak group $SU(2)$ must be extended because it is otherwise impossible to have different vacuum expectation values for the weak and the flavor symmetries by means of fhiggs fields in the fundamental representations. The groups $[U(1)_X]^2$ are two copies of an extra gauge symmetry we need in order to separate standard fermions from the exotic fermions the model requires because of the enlarged $SU(3)_W$ symmetry that turns the fermion weak doublets into triplets.

The breaking of $SU(10)$ into $SO(10)$ also breaks the subgroups
$[SU(3) \times U(1)]_W^2\times\,[U(1)_X]^2 $ and only a diagonal combination
survives. On the contrary, the flavor symmetry $U(1)_F$ survives the breaking and we eventually have that
 \bea
  U(1)_F \times [SU(3)
\times U(1)]_W^2\times [U(1)_X]^2 \nn \\ 
\rightarrow U(1)_F \times [SU(3) \times U(1)]_W \times U(1)_X
\, .
\eea
The breaking in the gauge sector $[SU(3) \times U(1)]_W^2 \times [U(1)_X]^2
\rightarrow[SU(3) \times  U(1)]_W \times U(1)_X $ leaves $10$ gauge bosons
massive. The remaining
$44$ Goldstone bosons can be labeled according to  representations of the $U(1)_F \times
[SU(3) \times U(1)]_W \times U(1)_X $ symmetry as two
complex fields $\Phi_1$ and
$\Phi_2$ that transform as  triplets of $[SU(3)]_W$, have the same  $U(1)_W$
and opposite $U(1)_F$ charges (they are not charged under the exotic gauge symmetry  $U(1)_X$); two complex fields $\Phi_3$ and
$\Phi_4$ that transform as  triplets of $[SU(3)]_W$, have the same  $U(1)_W$ (equal to that of $\Phi_{1,2}$)
and opposite $U(1)_X$ charge (they are not charged under the flavor symmetry  $U(1)_F$);
a sextet of complex fields $z_{ij}$; and four
complex fields $s$, $s_1$, $s_2$, and
$s_3$, which are flavor singlets.

All these fields are still Goldstone bosons with no potential; their potential arises after the explicit breaking of the symmetry to which we now turn.

 \textbf{Collective symmetry  breaking.}
 The effective lagrangian of the pseudo-Goldstone
bosons must contain terms that explicitly  break the $SU(10)$ global
symmetry. These terms provide masses of the order of the scale $f$ for the $s,\, s_i$ and $z_{ij}$ fields. However, each term separately preserve  enough symmetry to keep
the flhiggs fields  $\Phi_i$  exact Goldstone  bosons.
Only the simultaneous action of two or more of the terms (\textit{collective breaking})
 turns them into pseudo-Goldstone with  a potential, even though there is still no mass term. Quadratic terms for the flhiggs will come from the coupling to right-handed neutrino, as we shall discuss presently.

The effective lagrangian is given by the kinetic term
\be
\mathcal{L}_0 = \frac{f^2}{2}\Tr (D^\mu \Sigma) (D_\mu \Sigma)^* \, , \label{kinetic}
\ee
the covariant derivative of which couples the pseudo-Goldstone bosons to the gauge fields $W^a _{i_{\mu}}$, $B_{i_{\mu}}$ and $X_{i_{\mu}}$ of the $SU(3)_{_W{i}}$ , $U(1)_{_W{i}}$ and $U(1)_{_X{i}}$
respectively. The index $i$ runs over the two copies of each group. We denote the gauge couplings of $SU(3)_{_W{i}}$ by $g_i$ and those of $U(1)_{_W{i}}$ by $g^\prime_i$.

The lagrangian in \eq{kinetic}  gives mass to the $z_{ij}$ and  $s_3$  fields. On the other hand,
each  term of index $i$ preserves a $SU(3)$ symmetry so that  only when taken together they can give a contribution to the potential of the flhiggs fields.

At this point the fields $s$, $s_1$ and $s_2$ are still massless. To give them a mass, we introduce
plaquette terms---terms made out of components of the $\Sigma$ field that preserve enough symmetry not to induce masses for the flhiggs fields. 

We therefore have  the effective lagrangian
 \bea
\mathcal{L}& =& \mathcal{L}_0 + a_1^2 f^2 \left| \Sigma_{4,4}\right|^2
+ a_2^2 f^2 \left| \Sigma_{8,8}\right|^2
 + a_3^2 f^2 \left| \Sigma_{4,9}\right|^2 \nn \\
&+&  a_4^2 f^2 \left| \Sigma_{8,10}\right|^2
+ a_5^2 f^2 \left| \Sigma_{4,10}\right|^2
+ a_6^2 f^2 \left| \Sigma_{8,9}\right|^2\,, \label{efl}
 \eea
where $a_{i}$ are  coefficients of $O(1)$. The relative signs of the plaquette terms are in principle arbitrary and presumably fixed by the ultraviolet  completion of the theory.

In  the breaking of $[SU(3) \times
U(1)]_W^2 \rightarrow [SU(3) \times U(1)]_W $ nine gauge bosons
become massive with masses of order $O(f)$ and gauge couplings $g = g_1 g_2/\sqrt{g_1^2+g_2^2}$, $g^\prime = g_1^\prime g_2^\prime/\sqrt{g_1^{\prime 2}+g_2^{\prime 2}}$,  respectively. We identify $g$ with the $SU(2)$ weak coupling of the standard model.
These  heavy gauge bosons---because of their mixing with those with lighter masses to be identified with the standard model gauge bosons---induce corrections on many observables that we know to be constrained by high-precision measurements. Their presence is the major constrain on the scale $f$ and, accordingly, the naturalness of the model, as discussed for the littlest-Higgs model in \cite{littlehiggs2}. 

The effective potential for the flhiggs fields is given by the tree-level contribution coming from the
plaquettes and the one-loop Coleman-Weinberg effective potential arising from the gauge interactions.  

After integrating out the massive states by means of their equations of motion,
the potential of the four pseudo-Goldstone bosons $\Phi_{i}$, the flhiggs, is made of only quartic terms:
 \bea
  \label{pot1}
\mathcal{V}_1 [\Phi_i ] &=&\lambda_1 (\Phi_1^{\dag}\Phi_1)(\Phi_2^{\dag}\Phi_2) +\lambda_2 (\Phi_3^{\dag}\Phi_3)(\Phi_4^{\dag}\Phi_4)\nn \\
&+& \lambda_3
(\Phi_1^{\dag}\Phi_1)(\Phi_3^{\dag}\Phi_3)+\lambda_4
(\Phi_2^{\dag}\Phi_2)(\Phi_4^{\dag}\Phi_4) \nn\\
&+& \lambda_5
(\Phi_1^{\dag}\Phi_1)(\Phi_4^{\dag}\Phi_4)+\lambda_6
(\Phi_2^{\dag}\Phi_2)(\Phi_3^{\dag}\Phi_3) \\
 &+&   \xi_1 |\Phi_1^{\dag}\Phi_2|^2+ \xi_2 |\Phi_3^{\dag}\Phi_4|^2 \nn \\
&+& \xi_3(\Phi_1^{\dag}\Phi_3)(\Phi_2^{\dag}\Phi_4) + \xi_4  (\Phi_1^{\dag} \Phi_4)(\Phi_2^{\dag}\Phi_3)\,,  \nn 
 \eea
 where the coefficients are given in terms of the gauge coupling $g_i,\, g^\prime_i$ and plaquette coefficients $a_i$.

 Quadratic and quartic terms
   \be
\mathcal{V}_2 [\Phi_i ] =  \sum \mu^2_i (\Phi_i^{\dag}\Phi_i) +  \sum \chi_i (\Phi_i^{\dag}\Phi_i)^2 \label{pot2}
\ee
   that are necessary to induce vacuum expectation values for the flhiggs fields are not generated in the bosonic sector discussed so far.
 In order to introduce them we  couple the pseudo-Goldstone bosons to right-handed neutrinos with masses at the scale $f$. This means that the flavor and electroweak symmetry breaking of the model is triggered by the right-handed neutrinos. This is done again along the lines of the little-Higgs collective symmetry breaking:
 to prevent  quadratically
 divergent mass term for $\Phi_i$---and thus render useless what done up to this point---the Yukawa lagrangian of the
 right-handed neutrinos sector is constructed by terms that taken separately leave invariant some subgroups of the
 approximate global symmetry $SU(10)$. In this way the  flhiggs bosons receive a mass
 term only from diagrams in which all the approximate global
 symmetries of the Yukawa lagrangian are broken. Because of this collective breaking,  the one-loop
 contributions to the flhiggs masses are only logarithmic
 divergent.

As an example, consider the Yukawa lagrangian for two pairs of right-handed neutrinos $\nu^{1c}_L$, $\nu^{2c}_L$, $\nu^{\prime 1}_R$ and $\nu^{\prime 2}_R$ and their interaction with one of the flhiggs fields $\Phi_i$. The relevant terms can be written as
\bea \label{yukright1}
\mathcal{L}_Y^{\nu_{R}}&=& \eta_1 f \Big[ \overline{\nu_L^{1c}} \nu^{\prime 1}_R
\Big(1-\frac{\Phi_i^{\dag}\Phi_i}{2f^2} + \cdots \Big) \nn \\
&+& \frac{i}{f}\overline{\nu_L^{1c}} \Phi_i^{\dag} 
\nu^{\prime 2}_R+ \cdots \Big] +  \eta_2 f \overline{\nu_L^{2c}} \nu^{\prime 2}_R\, .
\eea
From \eq{yukright1} we see that the one-loop quadratically divergent contribution to the mass of $\Phi_i$ cancels out and a mass term can only come from diagrams with simultaneous insertions of both terms, that linear in $\Phi_i$ and proportional to $\eta_1$ and the mass term proportional to $\eta_2$. 

The complete right-handed neutrino sector is given by twelve 10-components multiplets.
The reason why we introduce so many fields is that we eventually want  different, and independent, mass terms $\mu_i^2$  to be induced in the effective potential by the right-handed neutrino sector.

After integrating out the right-handed neutrinos, we obtain  the divergent one-loop contributions to the
pseudo-Goldstone bosons masses:
\be
 \label{massephi}
\mu^2_i \simeq \eta_k^2\eta^2_j  
\frac{f^2}{(4\pi)^2} \log
\frac{\Lambda^2}{M_\eta^2}  \,,
\ee 
where in the logarithm we have generically
indicated the mass of right handed neutrinos with $M_\eta \simeq
\eta f$.

 From \eq{yukright1}, we can also estimate the one-loop
divergent contributions to the quartic terms  in the effective potential  of \eq{pot2} coming
 from the right-handed neutrino sector:
 \be
\chi_{i} \simeq \eta_{j}^4 \,\frac{f^2}{(4\pi)^2} \log
\frac{\Lambda^2}{M_\eta^2}  \,,
 \ee
where the coefficients $\eta_j$ need not be different as in \eq{massephi}.

The effective potential for the pseudo Goldstone bosons is therefore made of the sum of eqs.\ (\ref{pot1})
 and (\ref{pot2}). We want to find  vacuum expectation values for the flhiggs fields $\Phi_i$ in this potential that breaks the symmetry $[SU(3)\times
 U(1)]_W\times U(1)_F\times U(1)_X$ down to the electric charge group $U(1)_Q$.  Such a  vacuum is  given  by the field configurations
 \begin{widetext}
 \be
 \label{vfl}
\vev{\Phi_{1}}= \left(\begin{array}{c}
  0 \\
 v_W /2\\
  v_{F}/2 \\
\end{array}\right)
\quad
\vev{\Phi_{2}}= \left(\begin{array}{c}
  0 \\
 v_W /2\\
 - v_{F}/2 \\
\end{array}\right)
\quad 
\vev{\Phi_{3,4}}= \left(\begin{array}{c}
  0 \\
 0\\
  v_{X}/2 \\
\end{array}\right) \, .
 \label{vacuum}
 \ee
\end{widetext}
The conditions to be satisfied, in order for \eq{vacuum} to be a  minimum yield an expressions for the vacua $v_F$ and $v_W$
as functions of the coefficients of the effective potential together with
 some conditions on the masses $\mu_i$. Notice that all the relationships discussed can only be approximate since the coupling of the scalar fields to the fermions eventually introduces small corrections.

 \textbf{Experimental constrains and lightest Higgs bosons.}
 After symmetry breaking, the model is described at low-energy by a set of gauge and scalar bosons.  
 The structure of the gauge boson sector is complicated by the mixing of the standard model gauge bosons to the new states we have introduced.  Our general strategy is to impose that the charged currents of the model coincide with those of the standard model.  This done, we are essentially left with the theory of the standard model with the addition of a massive neutral gauge boson $Z^\prime$ and we must check that its presence affects  all parameters  within the experimental bounds. 
 
A more complete analysis would require that also the corrections arising from the effective operators induced by the heavy gauge bosons with masses $O(f)$  be included. They  enter  to $O(v_W^2/f^2)$ and force the scale $f$ to be above 4 TeV~\cite{littlehiggs2} in most little-Higgs models. Notice that the overall fit of these  corrections against the experimental electroweak data can in principle be improved by the presence in the flhiggs model of  additional parameters  thus lowering the scale $f$ with respect to that found in the case of the littlest Higgs model. 
We are more interested in the consistency of the framework than in its detailed realization and therefore we neglect, in this Letter, these $O(v^2_W/f^2)$ corrections. 
\cite{littlehiggs2}. Also notice ref.~\cite{Cheng} for a simple way, which can easily be applied to our model, to lessen the bound in~\cite{littlehiggs2}. 

The mass of the $Z^\prime$ gauge boson is bounded by data on Drell-Yan production (with subsequent decay into charged leptons) in $p\bar p$ scattering to be larger than 690 GeV~\cite{Abe}. 
We require that deviation in $\rho$ and the Weinberg angle be within $10^{-3}$ while the deviations in the neutral current coefficients to be less than $10^{-2}$. This choice put these deviations in the  tree-level parameters in the ball park of standard-model radiative corrections. 

The importance of these constrains resides in their fixing the values of the free parameters $v_F$ and $g^\prime$. 
The bound on the $\rho$ parameter essentially fixes the gauge coupling
$g^\prime \simeq 0.06$.
 Once $g^\prime$ has been fixed to this value, the bound  on the mass of $Z^\prime$ requires
 \be
v_F \gtap 1260 \quad \text{TeV}\, . \label{boundF}
\ee 
 We would like to have $v_F$ as close to $v_W=246$ GeV as possible but the phenomenological constrains force it to a higher scale that, in what follows, we  take at its lower bound. 
  
We now turn to the scalar  sector of the model.
 The number of scalar bosons can readily be computed:
 the number of degrees of freedom of  $4$ complex triplets is $24$. Of these $9$ are eaten by the gauge fields, while $1$---the would-be Goldstone boson of the spontaneous breaking of the $U(1)_F$ global symmetry---is eliminated, after introducing the fermions in the model,  by an anomaly. Therefore, the scalar sector contains    12 scalar bosons, ten of which are neutral, two charged. 
 
 For arbitrary coefficients of the potential we lack an analytic result for all their masses. Their values depend on $v_F$ and $g^\prime$ and, after having fixed them, it is a  function of the parameters of the potential. To obtain an estimate of these masses, we vary the numerical value of the coefficients in the potential by a Gaussian distribution around the natural value 1 with a spread of 20\% (that is $\sigma=0.2$). This procedure gives us average values of these masses with a conservative error and we can consider the result the natural prediction of the model.  For each solution we verify that all bounds on flavor changing neutral currents are satisfied. This is possible at such a low energy scale because the  relevant effective operators induced by the exchange of the flavor-charged flhiggs fields are suppressed by powers of the fermion masses over $f$~\cite{littleflavons}.
 
 The lightest neutral scalar boson (what would be called the Higgs boson in the standard model) turns out to have a mass $m_{h^0} = 317 \pm 80 $ GeV. This  a rather heavy---that is, larger than 200 GeV---Higgs mass is due to the value of $v_F$  we were forced to take in order to satisfy the bound in \eq{boundF}.  It is a value that  only partially overlaps with the 95\% CL of the overall fit of the  electroweak precison data that gives $m_{h^0} < 237$ GeV~\cite{LEP}. It  provides us with  a very testable prediction of the model.

Above the lightest scalar, the other scalar masses are spread, the heaviest of them reaching above $f$. The  lightest charged Higgs bosons has a mass $m_{h^\pm} = 560\pm 192$ GeV.

What is  the overall fine-tuning in  the model with the mass (given above) for the Higgs boson  and $f\simeq 3 $ TeV? It is around 10\% when we include the fine tuning in the parameters of the potential in \eqs{pot1}{pot2} required by the split between $v_F$ and $v_W$ and the $O( 3 \lambda_t f^2/(8 \pi^2) \log \Lambda^2/f^2)$ corrections to the  Higgs mass coming from the top-like extra fermions loop. Therefore, while there is still a certain amount of fine tuning, this is substantially less than in the standard model with a light Higgs boson.

\textbf{Fermion mass textures.}
The coupling of the flhiggs to the fermions gives rise to mass matrices  with well-defined textures generated by the flavor charge assignment and the powers of parameter $k=v_F^2/f^2$ that accordingly enter the Yukawa lagrangian~\cite{noi}.  In terms of this texture parameter, we can write the mass matrix for the neutrinos, after the see-saw originated by integrating out the heavy right-handed states,  to  $O(k)$ as
\be 
M^{(\nu)} =
\left(\begin{array}{ccc} 0 & 1 & O(k)\cr 1 & 0 & O(k)
\cr O(k) & O(k)  & 1
\end{array}\right) \, ,
\ee
 and
\be
 M^{(l)} =\left(\begin{array}{ccc} 0 &
0 & 0\cr O(k) & O(k) & 1\cr 0 &1& O(k)
\end{array}\right)\,, 
\ee
for the charged leptons. Similarly, we find to $O(k^2)$ for the up quarks
\be 
M^{(u)} =
\left(\begin{array}{ccc} 
0 & 0 & 0\cr 
0& 0 & O(k)\cr 
0 & O(k^2) & 1
\end{array}\right)
\ee
and
\be
M^{(d)} =\left(\begin{array}{ccc} 
0 & 0 & O(k^2)\cr 
0 & O(k^2) & 1 \cr 
0 &O(k^2)& 1
\end{array}\right)\
\ee
for the down quarks.
For $k$ of the order of the Cabibbo angle (and therefore $f\simeq 3$ TeV), these mass matrices reproduce--- for assigned values of the Yukawa coefficients of $O(1)$---in a satisfactory manner the experimental data on mass hierarchies and mixing angles. Notice that, on the other hand, the problem of the absolute value of the neutrino Yukawa couplings with respect to the others is only partially addressed by the low-energy see-saw taking place in the model.

The flhiggs model contains also a number of exotic fermionic states with masse at the TeV scale coming from the extended $SU(3)_W$ symmetry and from the implementation of the collective-breaking mechanism in the fermion sector.
Even though they are only weakly coupled to the standard fermions, they can  be used as further experimental signatures.


\textbf{Acknowledgments.}
The authors wish to acknowledge with thanks the
partial support of  the European TMR Networks HPRN-CT-2000-00148
and 00152 and the Benasque Center for Science for the hospitality during the completion of this  work.



\begin{thebibliography}{99}



\bibitem{littlehiggs}
N.~Arkani-Hamed, A.~G.~Cohen, E.~Katz and A.~E.~Nelson,
JHEP {\bf 0207}, 034 (2002)
[arXiv:hep-ph/0206021];
I.~Low, W.~Skiba and D.~Smith,
Phys.\ Rev.\ D {\bf 66}, 072001 (2002)
[arXiv:hep-ph/0207243];
D.~E.~Kaplan and M.~Schmaltz,
JHEP {\bf 0310}, 039 (2003)
[arXiv:hep-ph/0302049].
S.~Chang and J.~G.~Wacker,
Phys.\ Rev.\ D {\bf 69}, 035002 (2004)
[arXiv:hep-ph/0303001].
W.~Skiba and J.~Terning,
Phys.\ Rev.\ D {\bf 68}, 075001 (2003)
[arXiv:hep-ph/0305302].
S.~Chang,
JHEP {\bf 0312}, 057 (2003)
[arXiv:hep-ph/0306034].

\bibitem{littleflavons}
F.~Bazzocchi, S.~Bertolini, M.~Fabbrichesi and M.~Piai,
Phys.\ Rev.\ D {\bf 68}, 096007 (2003)
[arXiv:hep-ph/0306184];
Phys.\ Rev.\ D {\bf 69}, 036002 (2004)
[arXiv:hep-ph/0309182];
F.~Bazzocchi,
arXiv:hep-ph/0401105.

\bibitem{flavormodels-old}

H.~Harari, H.~Haut and J.~Weyers,
Phys.\ Lett.\ B {\bf 78}, 459 (1978).
C.~D.~Froggatt and H.~B.~Nielsen,
Nucl.\ Phys.\ B {\bf 147}, 277 (1979).
T.~Maehara and T.~Yanagida,
Prog.\ Theor.\ Phys.\  {\bf 61}, 1434 (1979).
G.~B.~Gelmini, J.~M.~Gerard, T.~Yanagida and G.~Zoupanos,
Phys.\ Lett.\ B {\bf 135}, 103 (1984).

\bibitem{littlehiggs2}
T.~Han, H.~E.~Logan, B.~McElrath and L.~T.~Wang,
Phys.\ Rev.\ D {\bf 67}, 095004 (2003)
[arXiv:hep-ph/0301040];
C.~Csaki, J.~Hubisz, G.~D.~Kribs, P.~Meade and J.~Terning,
Phys.\ Rev.\ D {\bf 67}, 115002 (2003)
[arXiv:hep-ph/0211124].

\bibitem{Cheng}
H.~C.~Cheng and I.~Low,
arXiv:hep-ph/0405243.

\bibitem{PDG}
S. Eidelman et al., Phys.\ Lett.\ {\bf B592} 1, (2004).

\bibitem{Abe}
F.~Abe {\it et al.}  [CDF Collaboration],
Phys.\ Rev.\ Lett.\  {\bf 79}, 2192 (1997).

\bibitem{LEP} The LEP Electroweak Working Group, {\tt http://lepewwg.web.cern.ch/LEPEWWG/plots\-/winter2004/}.

\bibitem{noi} F. Bazzocchi and M. Fabbrichesi, to appear.

\end{thebibliography}
 \end{document}